# Facile synthesis of micro-flower NiCo$_2$O$_4$ assembled by nanosheets efficient for electrocatalysis of water

Yujie Wang[a], Yan Duan[a, b], Yuwen Chen[b], Man Zhang[c], Yuchen Wang[b], Bin Liu[b], Xiaodie Zhang[b], Yutong Zhang[a, b], Kai Yan[b, d], *

[a] *Guangzhou Key Laboratory of Environmental Catalysis and Pollution Control, School of Environmental Science and Engineering, Institute of Environmental Health and Pollution Control, Guangdong University of Technology, Guangzhou 510006, China*
[b] *Guangdong Provincial Key Laboratory of Environmental Pollution Control and Remediation Technology, School of Environmental Science and Engineering, Sun Yat-sen University, Guangzhou 510275, China*
[c] *Key Laboratory of General Chemistry of the National Ethnic Affairs Commission, School of Chemistry and Environment, Southwest Minzu University, Chengdu 610041, China*
[d] *Guangdong Laboratory for Lingnan Modern Agriculture, South China Agricultural University, Guangzhou 510642, China*

ABSTRACT

Effective regulation of the morphology of transition metal spinel structures is crucial for creating efficient and stable bifunctional catalysts for electrocatalysis of water. In this work, micro-flower NiCo$_2$O$_4$ (F-NCO) assembled by nanosheets via a chemical template method for the simultaneous promotion of hydrogen evolution reaction (HER) and oxygen evolution reaction (OER). Electronic microscope analysis revealed that the thickness of the F-NCO catalyst was only 2.7% of that of the NiCo$_2$O$_4$ bulk (B-NCO), and this ultrathin lamellar structure was conducive to further exposure of the active site and improved reaction kinetics. The F-NCO catalyst exhibited superior HER and OER performance ($\eta_{10}$ = 236 and 310 mV) and robust long-term stability over the B-NCO catalyst in 1.0 M KOH, with a 2.68-fold and 4.16-fold increase in active surface area and a 0.42-fold and 0.61-fold decrease in charge transfer resistance values, respectively. Additionally, F-NCO/CFP$^{(+)}$||F-NCO/CFP$^{(-)}$ electrolyser was capable of producing 10 mA cm$^{-2}$ of electrocatalytic water decomposition at a low cell voltage of only 1.74 V, which is much better than the B-NCO/CFP$^{(+)}$||B-NCO/CFP$^{(-)}$ electrolyser. In a 12-hour stability test, the potentiometric time dropped by only 30 mV, which was less than that of the B-NCO electrode (70 mV). This micro-flower-structured electrode has remarkable electrocatalytic property and long-term durability, providing a novel insight for characterizing cost-effective and high-performance bifunctional electrocatalysts.

For the purpose of alleviating the growing environmental pollution and energy constraints, the quest to develop economically efficient and sustainable clean energy sources has attracted tremendous attention [1-3]. Hydrogen, as a clean energy with high natural abundance, high energy density, and harmless combustion product, stands out among a host of alternatives [4-6]. Compared with the unsustainable traditional fossil energy reforming and industrial by-product gas processes, hydrogen production from electrocatalysis of water can effectively avoid secondary pollution [7-9]. Generally, the electrocatalytic water decomposition process includes the two half-reactions: the cathodic hydrogen evolution reaction (HER) and the anodic oxygen evolution reaction (OER), with a theoretical equilibrium potential which compared with the reversible hydrogen electrode (RHE) for is 1.23 V, which [10-13]. The key to achieve efficient electrocatalysis of water lies in an ability to overcome the sluggish kinetic four-electron process (2H$_2$O → 2H$_2$ + O$_2$) and the high thermodynamic energy barriers (1.8 eV) [14, 15]. Up to date, noble metal catalysts (e.g., metallic Pt and Ir/Ru-based oxides) were extensively served as ideal electrocatalysts for accelerating HER or OER reactions by virtue of their empty d-orbitals and low inter-energy spacing [16]. Nevertheless, it is restrained by high price and rarity, which result in low commercial value for its practical application [17-19].

Transition metal spinel-structured catalysts (TMS) have been widely investigated for their elevated transition metal ion reactivity, electrochemical stability and unique electronic structure [18, 20, 21]. Since the binding capacity of active sites to reactants/intermediates takes an essential factor in OER and HER activities, the optimization of the catalyst structure/morphology is vital to the water electrocatalytic process [5, 22, 23]. The theoretical potentials of Ni and Co-based spinel for oxygen precipitation activity are comparable to those of noble metals, implying the ability to boost the four-electron transfer from the anodic OER during the electrocatalysis of water process [24, 25]. Compared to Ru-based catalysts, Ni- and Co-based oxide catalysts display optimal Cat-OH adsorption strength during hydrolysis dissociation (Volmer step: H$_2$O + e$^-$ + Cat → Cat-H$_{ads}$ + OH$^-$, as

---

* Corresponding author.
*E-mail address*: yank9@mail.sysu.edu.cn (K. Yan).

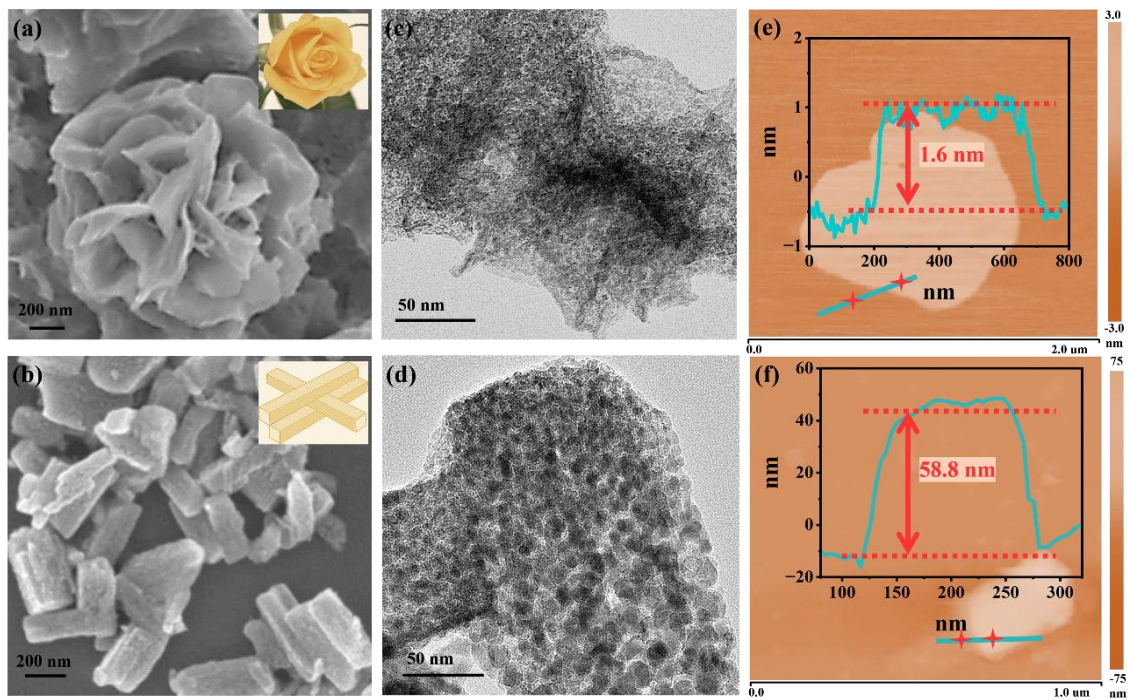

**Fig. 1.** Characterizations of crystallographic structures and microstructures: (a) SEM image, (b) TEM image and (c) AFM image of F-NCO. (d) SEM image, (e) TEM image and (f) AFM image of B-NCO.

Cat means the catalyst and $H_{ads}$ means the adsorbed H) [26]. $NiCo_2O_4$, a representative TMS, has the anti-spinel structure consisting of face-centered cubic lattices alternately stacked by oxygen octahedrons ($MO_6$) and oxygen tetrahedrons ($MO_4$) [27]. The cooperative of metal Ni and Co stabilizes the lattice distortion caused by the high spin state ($t_{2g}^4 e_g^2$) of single metal $Co^{3+}$, and lower the Jahn-Teller strain energy [28], thereby maintaining the stability of $NiCo_2O_4$ spinel [29, 30]. Moreover, the presence of mixed metals and redox pairs (Co 3p/Co 2p and Ni 3p/Ni 2p) not only reduces the activation energy of $NiCo_2O_4$ resistive transfer, but also promotes the interaction with water molecules to form M-O bonds, giving rise to reaction process, the problem of dramatic volume expansion/contraction can reduce the contact between particles of the electrode material, which in turn presents great difficulties in terms of cycling stability [26, 31]. It is expected to design appropriate morphology of $NiCo_2O_4$, which can achieve both high electrocatalytic activity and stability.

Herein, we successfully fabricated micro-flower $NiCo_2O_4$ (F-NCO) assembled by nanosheets. Atomic force microscopy revealed that the thickness of the F-NCO structure was only 1.6 nm in comparison with the $NiCo_2O_4$ bulk (B-NCO) ~ 58.8 nm, leading to facilitate the charge-transferring ability in the electrocatalysis of water process. Meanwhile, in the two half-reactions of HER and OER processes, the electrochemically active area of F-NCO was 2.68 and 4.16 folds higher than those of B-NCO, the Tafel slopes of F-NCO (9 mF $cm^{-2}$) were lower than those of B-NCO (40 mF $cm^{-2}$), respectively, and the charge transport resistances were reduced by 1.34 and 0.86 Ω, which reveals that F-NCO has a much faster reaction kinetics during the electrocatalysis of water process. In two-electrode measurements, the voltage was 1.74 V of F-NCO/CFP[(+)]||F-NCO/CFP[(-)] electrolyser with a current density of 10 mA $cm^{-2}$, which was lower than the B-NCO/CFP[(+)]||B-NCO/CFP[(-)] electrolyser (2.07 V). The stability analysis over a 10-hour period showed only the change of 30 mV in voltage, less than the 70 mV in B-NCO, which proving that F-NCO had a better electrochemical stability. Therefore, this work provides an ideal strategy by merely modifying the structural morphology of spinel to develop efficient and durable catalysts for electrocatalysis of water.

The synthesis process was carried out as described in Fig. S1, X-ray diffraction (XRD) measurements were firstly applied to verify the crystal structure of the samples. As displayed in Fig. S2, both F-NCO and B-NCO exhibited the typical $NiCo_2O_4$ configuration, and the peaks appearing at 31.1°, 36.7°, 44.5°, 59.0°, 64.7° correspond to the (220), (311), (400), (511), (440) crystal plane, respectively. The morphology and microstructure of F-NCO and B-NCO catalysts were characterized by scanning electron microscopy (SEM) and transmission electron microscopy (TEM). Fig. 1a, S3a and Fig. 1b, S3b displayed the SEM images of F-NCO and B-NCO catalysts, respectively. Compared with the B-NCO, F-NCO exhibited the micro-flower morphology self-assembled by nanosheets, which was favorable to provide sufficient active adsorption sites during the reaction. TEM images of F-NCO and B-NCO also evidenced the formation of such structures (As shown in Fig. 1c, S4a and Fig. 1d, S4b). The thickness of F-NCO and B-NCO was measured via atomic force microscopy (AFM), respectively. As displayed in Fig. 1e and 1f, the thickness of F-NCO was only 1.6 nm, and that of B-NCO was 58.8 nm. Three-dimensional peak force model images (Fig. S5) also reflected the presence of sheet and bulk structure, and revealed differences in the thickness of the F-NCO and B-NCO catalysts more visually. This characteristic structure of spontaneous assembly is favored for the exposure of metal active sites and boosted the charge transfer across the material and the s electrolyte.

The chemical valence of the elements in F-NCO was investigated through X-ray photoelectron spectroscopy (XPS). As displayed in Fig. 2a, the full spectrum provided evidence for the occurrence of Ni, Co and O elements in F- NCO. In Fig. 2b, the fitting of the Ni 2p orbital can obtain two orbital spin peaks in 856.5 and 874.0 eV, respectively, accompanying a pair of satellite peaks. In the Ni $2p_{3/2}$ level, $Ni^{2+}$ and $Ni^{3+}$ were related to the peaks of 854.5 and 856.9 eV, respectively, and the content of $Ni^{3+}$ was much higher than that of $Ni^{2+}$. This provided F-NCO with sufficient catalytic oxidation activity [32]. Similarly, the peaks at 780.1 and 782.1 eV in Fig. 2c illustrated that the ratio of $Co^{2+}$ and $Co^{3+}$ in Co $2p_{3/2}$ was in a balanced state. In Fig. 2d, the O 1s spectrum can be aligned to three peaks for lattice oxygen (O-M) ~530.5, surface-OH ~531.5 and adsorbed $H_2O$ ~532.8 eV. As shown in Fig. S6, metal Ni tended to exist in the electronic configuration of high spin $Ni^{3+}$ (HS, $t_{2g}^6 e_g^2$) and Co ions occupied tetrahedra and octahedra tend to favor the electronic configurations of HS $Co^{2+}$ ($t_2^3 e^4$) and low spin $Co^{3+}$ (LS, $t_{2g}^6 e_g^0$) in F-NCO [28], respectively. Metal orbitals at lower positions such as the e and $t_{2g}$ orbitals were completely dominated and not able to participate in bonding. The interaction between the high-position M-d orbitals (i.e., $t_2$ orbitals and $e_g$ orbitals) and the O-p orbitals ($p_x$, $p_y$, and $p_z$) token a real role in the activation of the water molecule [18]. This low electron density arrangement in F-NCO catalyst promoted the adsorption of negatively charged $OH^-$, thereby accelerating the reaction kinetics.

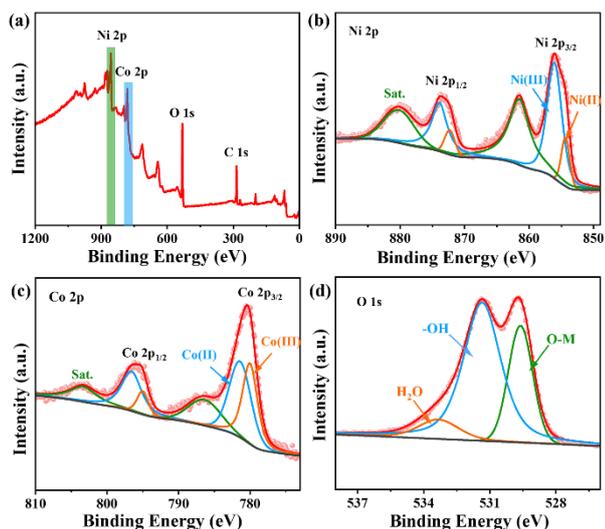

**Fig. 2.** Internal atomic structure analysis of F-NCO: (a) XPS spectra, (b) Ni 2p orbitals, (c) Co 2p orbitals, (d) O 1s orbitals.

A series of electrochemical tests were accomplished in 1 M KOH by using the conventional three-electrode system to evaluate the aqueous electrocatalytic properties of the F-NCO and B-NCO electrodes. The LSV polarization curves of F-NCO, B-NCO, $Co_3O_4$ and 20% Pt/C electrode in argon-saturated electrolyte were displayed in Fig. 3a. It can be noticed that the 20% Pt/C electrode showed the best HER performance. Fig. 3b demonstrated the overpotential at the current density of -10 mA cm$^{-2}$ was 68 mV, which was generally agreement with previous research results, while the overpotential of F-NCO electrode (236 mV) was inferior to that of B-NCO (283 mV) and $Co_3O_4$ (242 mV), suggesting that the kinetics of F-NCO electrode in HER was more favorable. Similarly, the F-NCO electrode exhibited lower Tafel slope than that of B-NCO and $Co_3O_4$, which indicated that the F-NCO electrode performs the best kinetics in the HER process and was in line with the Volmer-Heyrovsky mechanism [1, 11]. These results suggested that the nanosheet structure of F-NCO was conducive to lowering the energy barrier of the Heyrovsky step, thereby hastening the dissociation of $H_2O$. The internal electrode kinetic process was revealed by employing EIS, and the high-frequency region presents semicircles of different diameters (Fig. 3c). The values of equivalent series resistance (Rs) and charge transfer resistance (Rct) were summarized in Table S2. The smaller semicircle diameter (Rct = 1.88 Ω) of the F-NCO electrode in comparison with B-NCO (Rct = 3.22 Ω) and $Co_3O_4$ (Rct = 10.60 Ω) implied that the F-NCO electrode possessed the best interfacial charge transfer efficiency and better catalytic reaction kinetics [33], which was in accordance to Tafel slope analysis. ECSA of the electrodes were characterized through $C_{dl}$ calculated from CV, and the $C_{dl}$ of HER was further calculated (Fig. S7 and S8), which was higher for F-NCO electrode ($C_{dl}$=5.0 mF cm$^{-2}$) than that of B-NCO ($C_{dl}$=1.2 mF cm$^{-2}$) and $Co_3O_4$ ($C_{dl}$=2.1 mF cm$^{-2}$). The corresponding ECSA was 125, 30 and 52.5 cm$^2$, respectively, indicating the highest quantity of exposed active sites per unit area of the F-NCO surface, the greatest electron transfer capacity, and the best HER activity [34]. Meanwhile, comparing with previous studies, this micro-flower-structured catalyst which self-assembled from nanosheets behaved with a great adsorption area without further modification during electrocatalysis, which was extremely favorable to promote charge transfer, which was coherent with the excellent EIS [29, 35].

Meanwhile, the durability of the catalysts was also a very important evaluation criterion. As displayed in Fig. 3d, the stability of F-NCO electrode for HER was measured by i-t curve and multiple CV. After continuous operation at a potential of -0.18 $V_{RHE}$ for 10 hours, the current density remained at 87.2%. Similarly, the potential shift before and after 2000-cycle CV scanning was only 34.1 mV at the current density of -10 mA cm$^{-2}$. This result exhibited the well-stable nature of F-NCO electrode in HER.

The OER polarization curve of F-NCO and B-NCO electrodes was obtained in an oxygen-saturated electrolyte. As displayed in Fig. 4a, the F-NCO electrode had lower onset potential and higher current density compared with B-NCO, $Co_3O_4$ and commercial $RuO_2$ electrodes, which revealed that F-NCO had the best OER properties. The overpotentials of samples reached to 10 mA cm$^{-2}$ presented in Fig. 4b, and the F-NCO electrode only required 310 mV, which was smaller than that of B-NCO (390 mV), $Co_3O_4$ (340 mV) and $RuO_2$ electrodes (370 mV). According to the Tafel slope which were derived from the LSV curves, the F-NCO electrode had the smaller Tafel slope in all of the samples, which was

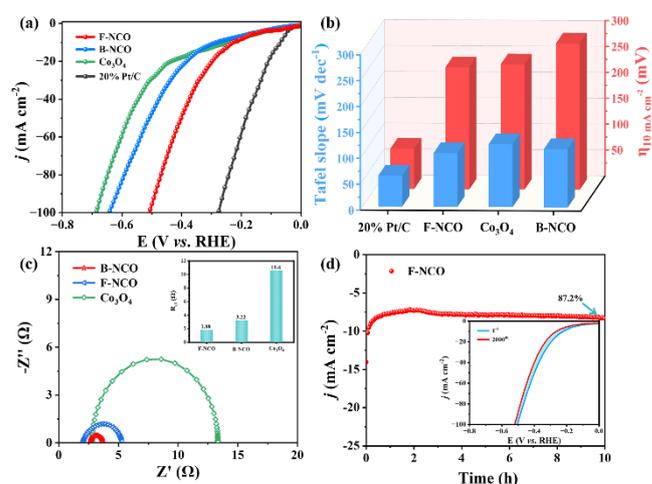

**Fig. 3.** Electrocatalytic activity for catalysis of HER (a) The polarization curves of F-NCO, B-NCO, $Co_3O_4$ and 20% Pt/C electrode. (b) Tafel slope and overpotential at current density of -10 mA cm$^{-2}$. (c) EIS of F-NCO and B-NCO, $Co_3O_4$, and the charge transfer resistance in the insets. (d) The CA curve of F-NCO at -0.18 $V_{RHE}$, and the comparison before and after 2000-cycle CV test for F-NCO electrode.

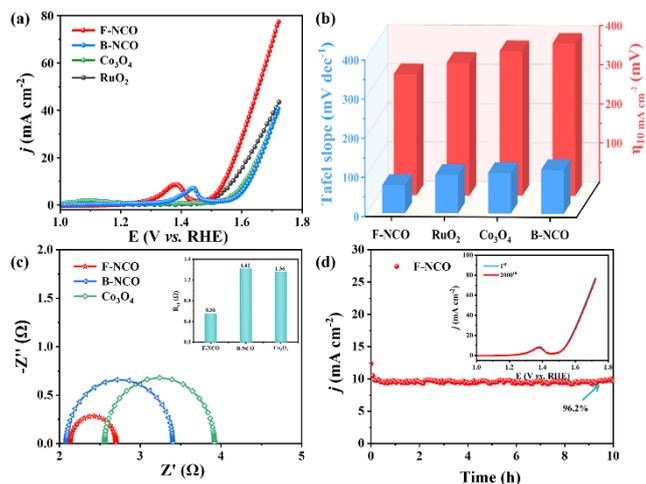

**Fig. 4.** Electrocatalytic activity for catalysis of OER. (a) The polarization curves of F-NCO, B-NCO, Co$_3$O$_4$ and RuO$_2$ electrode. (b) Tafel slope insets. (c) EIS of F-NCO and B-NCO, Co$_3$O$_4$, and the charge transfer resistance in the insets. (d) The CA curve of F-NCO at 1.52 V$_{RHE}$, and the comparison before and after 2000-cycle CV test for F-NCO electrode.

72 mV dec$^{-1}$, signifying the kinetics were optimized in OER process. Faster current density gains and smaller overpotential changes led to better electrocatalytic activity, which was consistent with the LSV results. Furthermore, EIS was employed to detect the reaction kinetics between the electrode and the interface. By fitting the Nyquist diagram with the classical equivalent circuit model (Fig. 4c), Table S3 summarizes the Rs and Rct values. According to these results, the F-NCO electrode manifested the smaller Rct (0.56 Ω) value among all the studied samples, which further indicated its superior charge transfer capability and outstanding electrical conductivity in terms of OER. Similarly, the C$_{dl}$ of OER was obtained at the same scan rate with HER following the CV curves in the non-Faraday intervals. As shown in Fig. S9 and S10, the C$_{dl}$ values of F-NCO electrode (4.3 mF cm$^{-2}$) was superior to that of the B-NCO (1.6 mF cm$^{-2}$) and Co$_3$O$_4$ electrodes (2.1 mF cm$^{-2}$), with the correlative ECSA was 107.5, 40 and 52.5 cm$^2$, respectively, which implied the greater density of active sites in the OER direction of F-NCO electrode.

Meanwhile, the durability in OER was evaluated by i-t curve and cyclic CV method. As shown in Fig. 4d, the 10 hours stability test was carried out at 1.52 V$_{RHE}$, and the i-t curve showed that the F-NCO electrode remained at 96.2%. As shown in inset of Fig. 4d, the potential shift was only 5.7 mV before and after the 2000-cycle CV assay at the current density of 10 mA cm$^{-2}$. The above properties and reliable stability in the OER process results indicated that the F-NCO electrode formed by simple morphology modulation manifested excellent.

Based on the remarkable HER and OER properties of F-NCO electrode, a two-electrode cell for electrocatalysis of water was assembled by using F-NCO electrodes as anode and cathode (Fig. 5a). From the electrochemical polarization curve (Fig. 5b), the F-NCO electrode required only 1.74 V to provide a current density of 10 mA cm$^{-2}$, significantly less than 1.93 V required for the B-NCO electrode. The inset graphs displayed the presence of hydrogen and oxygen of F-NCO electrode during the electrocatalysis of water process, confirming the excellent hydrogen/oxygen production capacity of F-NCO catalyst.

Additionally, it was found that the required potential of the F-NCO electrode increased by only 30 mV after being maintained 10 hours at 10 mA cm$^{-2}$, while this incremental trend was smaller than the 70 mV of B-NCO (Fig. 5c). Importantly, the LSV polarization curves to the constructed F-NCO electrode system did not observe obvious attenuation before and after 2000-cycle of CV test, demonstrating its long-term cyclic stability (Fig. S11). Meanwhile, XRD analysis of the reacted F-NCO electrode (As displayed in Fig. S12) revealed that the crystalline structure of the NiCo$_2$O$_4$ phase remained stable without the appearance of other phase products, which demonstrated that the F-NCO electrode had favorable stability during electrocatalysis of water process. The above results strongly proved that the synergistic effect between transition metals with high entropy structure characteristics, electronegativity and small atomic size difference was beneficial to show high efficiency in electrocatalytic water splitting, which was attributed to the formed simple solid solution during the electrocatalysis of water process. The F-NCO with sheet structure can provide sufficient adsorption sites and facilitate smooth charge transfer, resulting in satisfactory electrocatalytic activity and durability.

In summary, we had successfully fabricated F-NCO catalysts with micro-flower structure assembled by nanosheets via the chemical template method. Compared with B-NCO catalysts synthesized by solid-state method, F-NCO catalysts exhibited superior HER and OER activities, which was related to its sheet structure. AFM analysis showed that the thickness of F-NCO catalysts was only 2.7% of that of B-NCO catalysts, which was convenient for the charge transfer and adsorption between the catalyst and water molecules. ECSA showed that F-NCO increased by 4.16 times, the ion diffusion path was shorter and the electrocatalytic rate of water increased by 1.35 times. In the two half-reactions, the overpotentials of F-NCO catalysts were HER ~ 236 mV and OER ~ 310 mV achieved10 mA cm$^{-2}$, respectively, which were superior to the B-NCO catalysts. Additionally, F-NCO/CFP$^{(+)}$||F-NCO/CFP$^{(-)}$ electrolyser also exhibits excellent

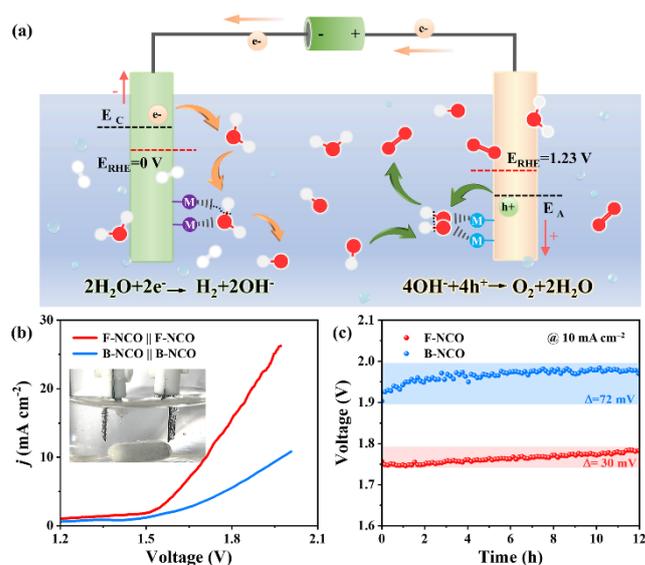

**Fig. 5** Analysis of electrochemical properties of electrocatalysis of water: (a) Schematic diagram of the electrocatalysis of water. (b) Polarization curves of F-NCO and B-NCO electrode and the electrocatalysis of water. (c) The 12-hour potential-time curve at 10 mA cm$^{-2}$ of F-NCO$^{(+)}$||$^{(-)}$F-NCO and B-NCO$^{(+)}$||$^{(-)}$B-NCO.

electrocatalytic performance, which was capable of generating 10 mA cm$^{-2}$ of electrocatalytic water decomposition at only 1.74 V. The F-NCO catalyst also showed remarkable stability over the 2000 cycles with almost negligible current fluctuations, the 12-hour potential-time curve at 10 mA cm$^{-2}$ showed only 30 mV potential fluctuation. This work provides an ideal way to obtain an efficient and stabilized bifunctional electrocatalysis of water catalyst by only modifying the morphology of spinel-based catalysts.

## Acknowledgements

This study funded by the National Key R&D Program of China (2023YFC3905804), the National Natural Science Foundation of China (22078374, 22378434), the Scientific and Technological Planning Project of Guangzhou (202206010145), and the Guangdong Provincial Key Laboratory of Environmental Pollution Control and Remediation Technology (2023B1212060016).

## References

[1] C. Hu, L. Zhang, J. Gong, Energy Environ. Sci. 12 (2019) 2620-2645.
[2] M. Batool, A. Hameed, M. Nadeem, Coord. Chem. Rev. 480 (2023) 215029.
[3] J. Wang, Y. Gao, H. Kong, et al., Chem. Soc. Rev. 49 (2020) 9154-9196.
[4] L. Chong, G. Gao, J. Wen, et al., Science 380 (2023) 609-616.
[5] Z. Yu, Y. Duan, X. Feng, et al., Adv. Mater. 33 (2021) e2007100.
[6] H. Luo, P. Yu, G. Li, et al., Nat. Rev. Phys. 4 (2022) 611-624.
[7] J. Qi, Y. Lin, D. Chen, et al., Angew. Chem. Int. Ed. 59 (2020) 8917-8921.
[8] X. Wang, Y. Qin, X. Peng, et al., Inorg. Chem. Front. 11 (2024) 957-969.
[9] Y.-Y. Li, X.-H. Li, Z.-X. An, et al., Chin. Chem. Lett. DOI: 10.1016/j.cclet.2024.109716
[10] J. Sun, H. Xue, Y. Zhang, et al., Nano Lett. 22 (2022) 3503-3511.
[11] H. Jin, J. Wang, D. Su, et al., J. Am. Chem. Soc. 137 (2015) 2688-2694.
[12] A. Li, S. Kong, C. Guo, et al., Nat. Catal. 5 (2022) 109-118.
[13] S. Iqbal, B. Safdar, I. Hussain, et al., Adv. Energy Mater. 13 (2023) 2203913.
[14] X. Yan, M. Xia, H. Liu, et al., Nat. Commun. 14 (2023) 1741.
[15] J. Xu, X. Li, Z. Ju, et al., Angew. Chem. Int. Ed. 58 (2018) 3032-3036.
[16] P. Kuang, Z. Ni, B. Zhu, et al., Adv. Mater. 35 (2023) 2303030.
[17] K. Yan, Y. Lu, W. Jin, ACS Sustain. Chem. Eng. 4 (2016) 5398-5403.
[18] Y. Huang, M. Li, F. Pan, et al., Carbon Energy 5 (2022) e279.
[19] J. Kibsgaard, T. Jaramillo, Angew. Chem. Int. Ed. 53 (2014) 14433-14437.
[20] L. Tao, K. Pang, L. Huang, et al., Catal. Lett. 153 (2022) 2642-2650.
[21] Z. Zhang, X. Li, H. Tang, et al., Chin. Chem. Lett. DOI: 10.1016/j.cclet.2024.109700
[22] Q. Wang, H. Wang, X. Cheng, et al., Mater. Today Energy 17 (2020) 100490.
[23] F. Chen, W. Peng, J. Zhou, et al., Inorg. Chem. Front. 11 (2024) 425-435.
[24] I. Man, H. Su, F. Calle, et al., ChemCatChem 3 (2011) 1159-1165.
[25] M. Zhang, D. Hu, Z. Xu, et al., J. Mater. Sci. Technol. 72 (2021) 172-179.
[26] D. Wang, Y. Chen, L. Fan, et al., Appl. Catal. B: Environ. 305 (2022) 121081.
[27] Y. Fu, C. Peng, D. Zha, et al., Electrochimi. Acta 271 (2018) 137-145.
[28] J. Wu, X. Wang, W. Zheng, et al., J. Am. Chem. Soc. 144 (2022) 19163-19172.
[29] L. Wang, C. Gu, X. Ge, et al., Adv. Mater. Interfaces 4 (2017) 1700481.
[30] L. Zhang, J. Peng, W. Zhang, et al., J. Power Sources 490 (2021) 229541.
[31] Y. Wang, Y. Liu, A. Ejaz, et al., Chin. Chem. Lett. 34 (2022) 107538.
[32] B. Liu, Z. Zheng, Y. Liu, et al., J. Energy Chem. 78 (2023) 412-421.
[33] B. Mao, P. Sun, Y. Jiang, et al., Angew. Chem. Int. Ed. 59 (2020) 15232-15237.
[34] Z. Xie, W. Wang, D. Ding, et al., J. Mater. Chem. A 8 (2020) 12169-12176.
[35] Depeng Zhao, Meizhen Dai, Hengqi Liu, et al., Adv. Mater. Interfaces 6 (2019) 1901308.